\documentclass[preprint,12pt]{elsarticle}
\usepackage{amsmath}
\usepackage{amssymb}
\usepackage{graphicx}
\usepackage{float}
\newcommand{\lgmode}[2]{LG$_{#1}^{#2}$}
\journal{~}
\begin{document}
\begin{frontmatter}
	\title{Confinement of ultracold atoms in a Laguerre-Gaussian laser beam created with diffractive optics}
	\author{Sharon A. Kennedy}
	\author{G.~W.~Biedermann} 
	\author{J.~Tom Farrar}
	\author{T.~G.~Akin}
	\author{S.~Krzyzewski}
	\author{E.~R.~I.~Abraham}
	\cortext[cor]{Corresponding author.  Tel.: + 1 405 325 2961; Fax: +1 405 325 7557.}
	\ead{abraham@nhn.ou.edu}
	\address{Homer L. Dodge Department of Physics and Astronomy, University of Oklahoma, 440 W. Brooks St. Norman, OK 73019, USA\corref{cor}}
	\begin{abstract}
		We report 2D confinement of $^{87}$Rb atoms in a Laguerre-Gaussian laser beam.
		Changing of the sign of the detuning from the atomic resonance dramatically alters the geometry of the confinement.
		With the laser detuned to the blue, the atoms are confined to the dark, central node of the Laguerre-Gaussian laser mode.
		This trapping method leads to low ac Stark shifts to the atomic levels.
		Alternatively, by detuning the laser to the red of the resonance, we confine atoms to the high intensity outer ring in a multiply-connected, toroidal configuration.
		We model the confined atoms to determine azimuthal intensity variations of the trapping laser, caused by slight misalignments of the Laguerre-Gaussian mode generating optics.
	\end{abstract}
	\begin{keyword}
		Laguerre-Gaussian mode \sep Toroid trap \sep Multiply-connected trap \sep Atom trap \sep Diffractive optics
	\end{keyword}
\end{frontmatter}
\date{\today}
A laser propagating in the Laguerre-Gaussian (\lgmode{p}{\ell}) transverse mode is a versatile tool in atomic, molecular, and optical (AMO) physics. 
The $e^{\imath\ell\phi}$ azimuthal winding phase gives rise to an intrinsic quantized orbital angular momentum of $\ell\hbar$ per photon \cite{abs92}.
The additional quantum number may allow multi-dimensional quantum computing and encryption~\cite{mtt07}.
Experiments have been proposed \cite{mzw97} and demonstrated \cite{wlb08} that orbital angular momentum can be coupled from the optical field to the atomic internal states, revealed as a vortex state in a Bose-Einstein condensed (BEC) gas.
The $p+1$ radial intensity nodes create cylindrically symmetric geometries, generating a manifold of multiply connected traps. \\
\indent Blue-detuned dipole traps attract atoms to regions of low intensity, where the atoms scatter fewer photons and experience smaller ac Stark shifts.  
Cold atoms trapped in this configuration may be used in an atomic clock, and other precision measurements.
Dark optical traps have confined large numbers of atoms using an \lgmode{0}{3} beam \cite{kts97}, an arrangement of blue detuned lasers \cite{nla95}, and in other complex laser modes \cite{cfk97,okd99,okd99e,okf00,kac01,iwd09}. 
A dark toroidal geometry trap was created using a superposition of \lgmode{p}{\ell} beams \cite{otb07}.
A blue-detuned dipole trap generated from a spatial light modulator \cite{xhw10} and two crossed \lgmode{p}{\ell} laser modes generated from a spatial phase plate \cite{lzi12} trapped single atoms with long coherence times.
For the \lgmode{0}{\ell} mode laser, the azimuthal winding phase ensures that the intensity necessarily goes to zero at the center of the laser beam, confining the atoms in the central region of the laser beam. \\ 
\indent Alternatively, red-detuned dipole traps confine atoms in the high intensity region of the trapping laser.
Degenerate gases excited into vortex states confined in toroidal geometry traps exhibit unique matter wave interference patterns \cite{tda01}. 
Multiply connected traps locally pin vortices in BECs, making them ideal for ring shaped BEC rotational gyroscopes \cite{tkd12}.  
While it is possible to create a multiply connected trap using a magnetic trap whose center is plugged by a blue detuned fundamental mode laser \cite{rac07}, these traps are species and state selective.
Optical traps circumvent this problem.
For a red-detuned \lgmode{0}{1} mode laser, the atoms are confined in a toroidal geometry. 
A theoretical analysis has concluded that a BEC whose initial conditions are similar to those found in standard traps can be loaded into an \lgmode{p}{\ell} mode \cite{wad00}, and a theoretical calculation of the transition of a  thermal gas to a BEC within the \lgmode{p}{\ell} laser mode itself has been done \cite{akd12}. \\
\indent Diffractive optics can transform, external to the laser cavity, the Gaussian output of a laser into \lgmode{p}{\ell} modes. 
A diffractive optic is a transparent optic where lithography techniques are used to etch microscopic structures on the surface.
These structures are designed such that the laser wavefront evolves into the desired form via Huygens' principle.
Two optics are necessary to control both the intensity and phase.
Diffractive optics can create high-order \lgmode{p}{\ell} modes~\cite{kks00}, with demonstrated mode purities much higher than those formed with other methods \cite{kst02}. 
The compact diffractive optics have proven advantageous in quantum information processing, where the large numerical aperture of these optics gather the largest fraction of the fluorescence emitted by an ion trapped on a chip~\cite{snj11,kmk11,bem11}.
Recently, diffractive optic elements have been used to create blue detuned bottle beam traps \cite{iis13}.\\
\indent We report 2D confinement of ultracold $^{87}$Rb atoms loaded from a magneto-optical trap (MOT) in both blue-detuned and red-detuned \lgmode{0}{1} (doughnut mode) laser beams.  
The atoms are confined to the central node of the \lgmode{0}{1} mode in the blue-detuned case, and the atoms are confined in the toroidal anti-node of the \lgmode{0}{1} for the red-detuned case.
We align our \lgmode{0}{1} mode vertically (along the direction of gravity), maintaining cylindrical symmetry.
Study of the atomic density distribution reveals asymmetries in the \lgmode{0}{1} mode.
This provides an \textit{in situ} measurement of the \lgmode{p}{\ell} mode purity and the possibility of more exotic confinement potentials. \\
\begin{figure}
	\centering
	\includegraphics[width=8cm]{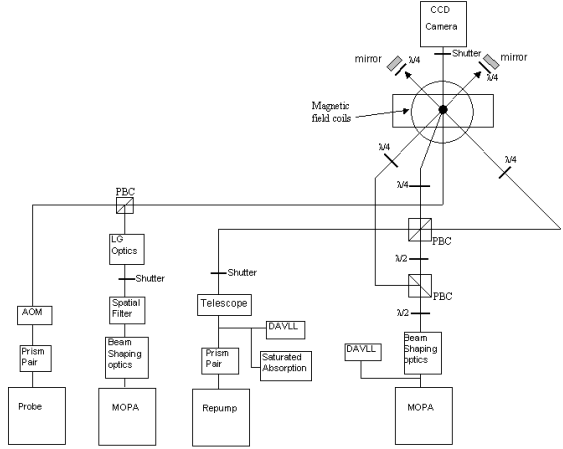}
	\caption{
		This figure is a conceptional diagram of our experimental setup.  
		Two of the MOT beams are directed at 45$^{\circ}$ relative to the cell.  
		The third beam, shown as a dot, goes through the face of the cell in the direction of the magnetic field.  
		The collinear probe and LG$^1_0$ laser beams are directed vertically through the center of the MOT and are then sent into a CCD camera.
		\label{experimental_figure}
	}
\end{figure}
\indent Figure \ref{experimental_figure} shows a schematic of the experiment.  
The trapping laser consists of a low-power external-cavity diode laser \cite{msw92} amplified by an SDL tapered amplifier in the Master Oscillator Power Amplifier (MOPA) configuration locked at a frequency to the red of the $| ^2S_{1/2},F=2\rangle\rightarrow | ^2P_{3/2},F'=3\rangle$ cycling transition in $^{87}$Rb using a Dichroic-Atomic-Vapor Laser Lock (DAVLL) \cite{mht07}.  
A series of three cylindrical lenses shape and expand the elliptical 300~mW output of this laser to a symmetrical Gaussian profile with an $1/e^{2}$ beam radius of 12.5~mm.  
The beam is sent through two polarizing beam splitter cubes (PBC) to create three beams that are directed through the cell along orthogonal axes and then retro-reflected.  
Three $\lambda/2$ retarding optics adjust the fraction transmitted and reflected by the PBCs, and thus the relative intensity of the three beams.  
Two of the beams are directed at 45$^\circ$ angles relative to the cell, while the third beam is directed along the axis of the anti-Helmholtz coils through the side of the cell. \\
\indent A 10~mW external cavity diode laser repumps the atoms that fall into the $|F=1\rangle$ hyperfine level back into the cycling transition.  
After reshaping the output with an anamorphic prism pair and expanding the beam to an $1/e^{2}$ beam radius of 12.5~mm, we inject the repump laser through the back of one of the PBCs so that it is collinear with the trapping beam.  
This laser is tuned to the $| ^2S_{1/2},F=1\rangle\rightarrow | ^2P_{3/2},F'=2\rangle$ transition.  
Originally, this laser was locked on resonance using saturated absorption spectroscopy.  
However, we find it easier to use the saturated absorption spectrometer \cite{msw92} to find the right transition, and lock the laser with a DAVLL.  
We then optimize the frequency of the repump laser by simply maximizing the number of trapped atoms in the MOT. \\
\indent The vapor-cell MOT is created from a Rb vapor in a 1.4~x~1.4~x~11~in rectangular glass cell that offers large optical access.  
This cell is pumped with a Varian Turbo-V~250~$\ell$/s vacuum pump roughed by a Varian DS~102 mechanical pump.  
Typical vacuum pressure is $10^{-8}$ to $10^{-7}$ Torr.  
At the center of the trap a magnetic field gradient of 20~G/cm is created by supplying 13~A through two 20-turn, 11~cm diameter coils placed 8~cm apart in an anti-Helmholtz configuration.  
The MOT regularly produces $\sim\!10^8$ atoms with  temperatures between 0.1-2.0~mK.  
The lower temperatures are achieved by adding three sets of Helmholtz coils that zero the magnetic field at the center of the MOT.\\
\indent To optically confine the cold atom sample, we overlap a laser in the \lgmode{p}{\ell} mode with the MOT.
The \lgmode{p}{\ell} modes have a radial electric field whose magnitude is proportional to the product of a Gaussian and an associated Laguerre polynomial $L^\ell_p(x)$, which gives the characteristic $p+1$ radial intensity nodes when $\ell>0$.  
For a planar wave front propagating along the $z$-axis, the magnitude of the electric field at $z=0$ is given by:
\begin{equation}
	\label{efieldmag}
		u^\ell_p(r)=\sqrt{{2P}\over{\pi w^2}}(-1)^p e^{-\imath\ell\phi}e^{-r^2/w^2}
			\left({\sqrt{2}r}\over{w}\right)^{|\ell|} L^\ell_p\left(2 r^2/w^2\right),
\end{equation}
where $P$ is the laser power and $w$ is the beam waist.  
The $e^{-\imath\ell\phi}$ term in Equation (\ref{efieldmag}) implies that there is a quantized azimuthal phase change of $2\pi\ell$ in the electric field.  
This results in an intensity node at the center of the beam and an angular momentum of $\ell\hbar$ per photon.  
A Gaussian beam occurs when $\ell=p=0$, whereas a donut beam occurs when $\ell>0$ and $p=0$.  
As $\ell$ increases, so does the orbital angular momentum, and thus so does the effective size of the central node.  
This can be seen from Equation \ref{efieldmag}, where the intensity has a functional form of $r^{2\ell}$ near $r=0$. \\
\indent Cold atoms can be confined to either the nodes or the antinodes of these beams by means of the optical dipole force potential, which in the approximation of a two-level system and in the limit of the detuning being large compared to the natural linewidth, has the form: 
\begin{equation}
	\label{potential}
	U(r)=\frac{\hbar\Gamma^2}{4\Delta}\frac{I(r)}{I_{\mathrm{sat}}}, 
\end{equation}
where $I(r)$ is the intensity distribution of the laser, $I_{\mathrm{sat}}$ is the saturation intensity, $\Delta$ is the detuning (the difference between the laser frequency and the transition frequency between the two states), and $\Gamma$ is the linewidth of the excited atomic state.
If the \lgmode{p}{\ell} laser is tuned above resonance (blue detuned, $\Delta>0$), then the atoms will be repelled from regions of high intensity and confined to the nodes of the \lgmode{p}{\ell} laser mode, whereas if the laser is tuned below resonance (red detuned, $\Delta<0$) then the atoms will be attracted to the anti-nodes.\\
\indent The \lgmode{0}{1} laser beam is created using diffractive optics developed in collaboration with the research department of Diffractive Optics Corporation.  
These optics are advantageous in that they offer an external cavity method of creating very pure, higher-order \lgmode{p}{\ell} laser modes.  
Creating high-order \lgmode{p}{\ell} beams with computer generated holograms has a maximum mode purity of 80\% intensity in the $p$ mode of interest \cite{ada98}. 
Our diffractive optics can generate an \lgmode{p}{\ell} beam with 100\% of the intensity in the desired mode \cite{kst02}.  
We use two optics to generate the desired \lgmode{p}{\ell} beam, one to control the intensity, and the other to control the phase.\\
\indent We spatially filter the 300~mW output of a second MOPA to obtain a more pure Gaussian beam.  
(Previously, for the highest mode purity we used a single mode optical fiber to filter the beam.) 
The resulting 100~mW output of the spatial filter is telescoped to a $1/e^{2}$ beam radius of 0.5~mm and sent through the two \lgmode{0}{1} optics.
We routinely get $\simeq 30$~mW of power in a pure \lgmode{0}{1} laser mode.  
The beam is then expanded by approximately a factor of four to increase the trapping volume.
This results in a radial trap frequency of 30~Hz, at a detuning of 2~GHz.
The beam is directed vertically through the center of the MOT. \\
\indent A weak probe laser is resonant with the $| ^2S_{1/2},F=1\rangle\rightarrow | ^2P_{3/2}F'=2\rangle$ transition, and propagates collinearly with the \lgmode{0}{1} beam.
The probe beam is shuttered using a NEOS N23080 Acoustic Optical Modulator, and is used to image the cloud of atoms onto a Pulnix TM-300NIR CCD camera. 
Because the intensity of the \lgmode{0}{1} beam saturates our camera, a Uniblitz LS6T2 shutter is placed in the \lgmode{0}{1} beam path and a second shutter placed in front of the camera, so the \lgmode{0}{1} beam never enters the camera.  
First, the \lgmode{0}{1} shutter is closed.  
After 0.1~ms, the camera shutter is opened, and the probe is flashed 1.9~ms later.  
The shutter in front of the camera does not always fully open for shorter delays. \\
\indent The cold atoms absorb the resonant light from the weak probe beam, and there is a reduction in intensity following Beer's law: $\ln(I_0/I_{\mathrm{out}}) = \sigma_0 n$.
Here, $I_0$ is the incident probe intensity, $I_{\mathrm{out}}$ is the probe intensity after passing through the atomic ensemble, $\sigma_0$ is the resonant photon scattering cross-section, and $n$ is the integrated column density.
In order to extract the atomic column density, we take a series of three CCD images.
The probe beam is first imaged in the presence of the trapped atoms, measuring $I_{\mathrm{out}}$.
Then, $I_0$ is measured by imaging the probe absent any confined atoms.
After blocking the probe laser, we expose the CCD chip and subtract this information ($I_{\mathrm{bg}}$) from the previous two images.
The column density at each CCD pixel is determined by:
\begin{equation}
	\label{column_density}
	n = \frac{1}{\sigma_0} \ln\left(\frac{I_0 - I_{\mathrm{bg}}}{I_{\mathrm{out}} - I_{\mathrm{bg}}}\right).
\end{equation}
The array of CCD pixels map out the column density in a plane transverse to the direction of the probe beam propagation. \\
\begin{figure}
	\centering
	 \includegraphics[width=8cm]{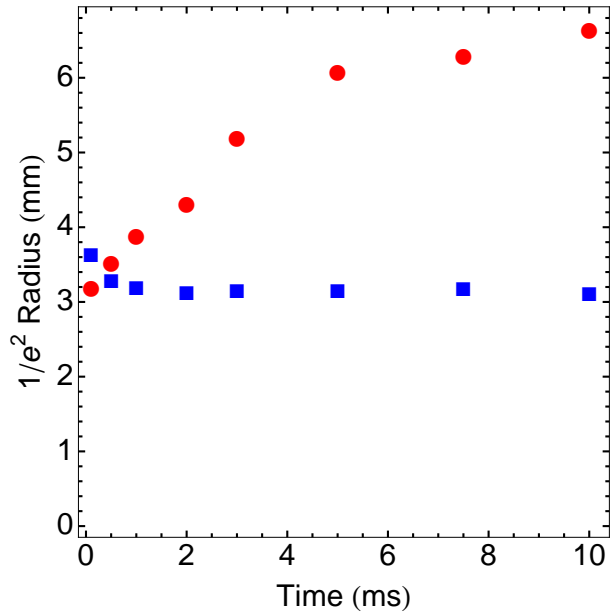}
	 \caption{
	 	(Color online) The size of the atom cloud as a function of expansion time.  
		The red circles represent the size of the atom cloud when there was no \lgmode{0}{1} beam present.  
		The blue squares represent the size of the cloud with an \lgmode{0}{1} laser beam present detuned to the blue of the atomic resonance. 
		With the \lgmode{0}{1} laser present, the size of the cloud remains approximately the same.  
		This implies that the atoms are indeed confined to the center node of the \lgmode{0}{1} laser mode.
	\label{blue}
	}
\end{figure}
\indent We confine $10^7~^{87}$Rb atoms in the central node of an \lgmode{0}{1} laser beam detuned 2~GHz to the blue of the $| ^2S_{1/2},F=1\rangle\rightarrow | ^2P_{3/2}\rangle$ resonance frequency.  
We first superimpose the \lgmode{0}{1} beam over the MOT.  
The repump laser is blocked, and the atoms are optically pumped into the $|F=1\rangle$ hyperfine level where they no longer interact with the MOT beams.  
The atoms are confined in 2D due to the repulsive force from the toroidal \lgmode{0}{1} beam surrounding the atoms.
To detect the atoms we block the \lgmode{0}{1} beam and probe the remaining atoms with the absorption imaging procedure described above.
We vary the delay between blocking the repump and probing the atoms.
If the atoms are confined to the central node of the \lgmode{0}{1} beam, then the size of the atom cloud will not change with longer delays.  
If there is no confinement, then the atom cloud will expand at a rate related to its temperature. \\
\indent The image from the CCD camera is fit to a Gaussian profile from which we can obtain the $1/e^2$ radius of the atom cloud.  
We find the density distribution is modelled accurately by a Gaussian.
Figure \ref{blue} shows the size of the atom cloud as a function of the delay between when the repump is blocked and the atoms are observed.  
The size of the atom cloud with and without the \lgmode{0}{1} beam are shown as blue squares and red circles respectively.  
When the \lgmode{0}{1} beam is superimposed over the MOT, the size of the atom cloud does not vary significantly as the delay time increased, indicating that the atoms are indeed confined to the central node of the \lgmode{0}{1} beam.  
When the \lgmode{0}{1} beam is not superimposed over the MOT, the atom cloud increases at a rate consistent with a temperature of 3~mK.
The discrepancy in the cloud sizes between the two data sets at $t=0$ is consistent with run-to-run variations. 
It is also possible that the repulsive interactions of the of the blue-detuned \lgmode{0}{1} mode with the atoms at the edge of the MOT distorts the initial distribution, making it larger than the MOT without the \lgmode{0}{1} beam.
The initial decrease in the optically confined cloud size is due to the loss of those atoms from the MOT not confined by the \lgmode{0}{1} mode.\\
\begin{figure}
	\centering
	\includegraphics[width=8cm]{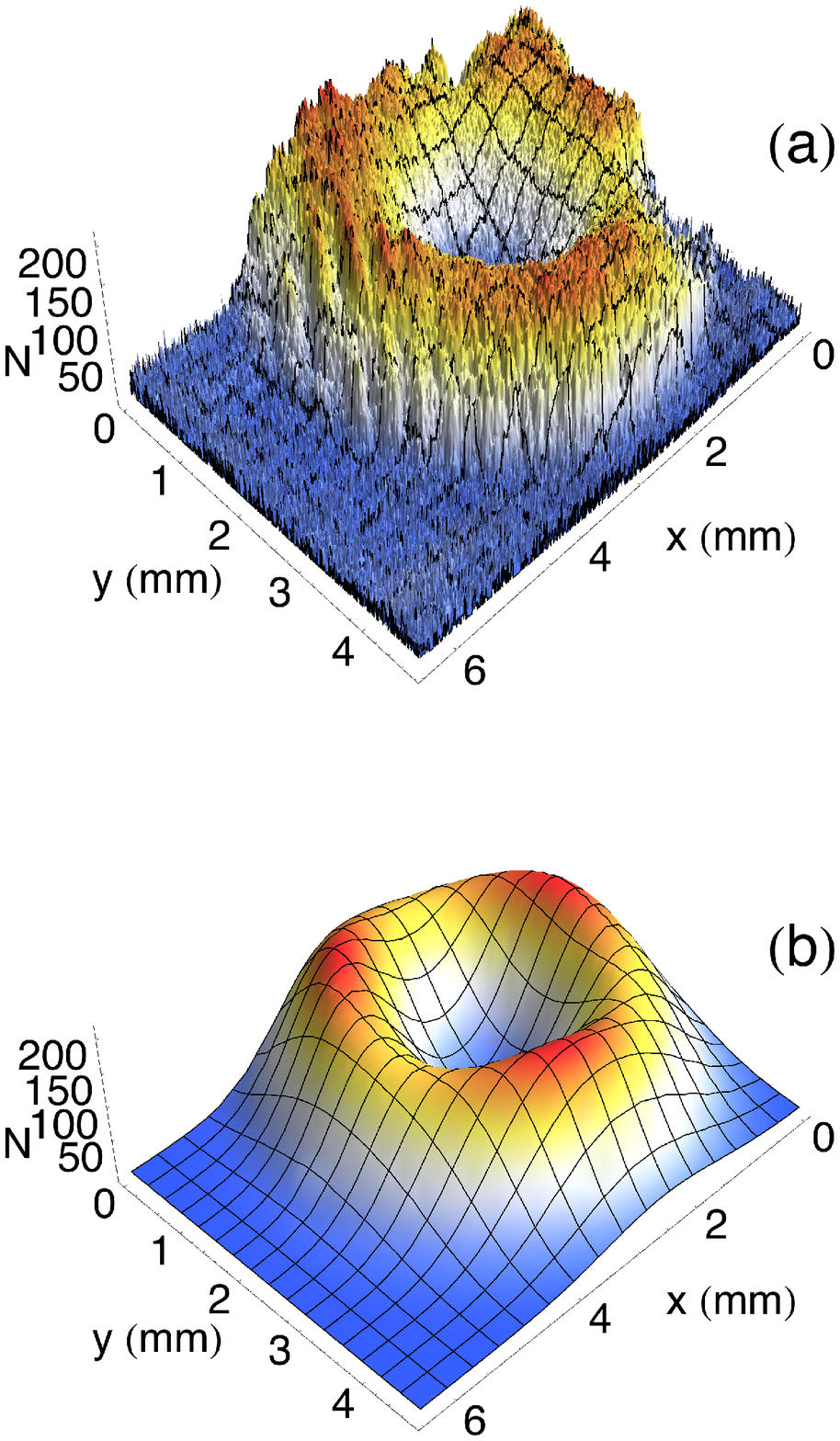}
	\caption{
		(Color online) (a) 3D CCD image of the number of atoms in the high-intensity region of our red-detuned LG$^1_0$ laser mode.  
		In this image the Gaussian shaped profile of the atoms still confined in the MOT, but not confined in the high-intensity area has been removed.
		(b) Fit of Equation~(\ref{density_azimuthal}) to the atom distribution in (a).  
		\label{redwoGaussian}
	}
\end{figure}
\indent Figure \ref{redwoGaussian}~(a) shows an absorption image of $10^4$ ultracold atoms trapped in a toroidal potential formed by a 30~mW \lgmode{0}{1} beam tuned 2~GHz to the red of the $| ^2S_{1/2},F=2\rangle\rightarrow{| ^2P_{3/2}\rangle}$ transition.  
To perform this experiment we superimpose the \lgmode{0}{1} beam over the MOT, and image the atoms as described above.
In this case, we do not turn off the MOT, so there are still atoms confined within the MOT that are not trapped in the \lgmode{0}{1} beam.  
We have subtracted these atoms from the absorption image to obtain Figure~\ref{redwoGaussian}~(a). \\
\indent The density distribution contained within the \lgmode{0}{1} is consistent with a Gaussian that is offset and rotated around the axis of symmetry.
However, there is an azimuthal asymmetry in the density distribution, from variations in the intensity of the \lgmode{0}{1} laser beam due to misalignment of the diffractive optics \cite{kst02}.  
This asymmetry in the intensity distribution manifests itself in the trap potential which causes a larger trap depth and more atoms in regions of higher intensity, shown in Figures~\ref{redwoGaussian}-\ref{azimuthal}. \\ 
\indent To model the system we assume the atoms obey Maxwell-Boltzmann statistics and the atomic density distribution is given by \cite{gwo00}:
\begin{equation}
	\label{density}
	n(r,\phi) = n_0 e^{-U(r,\phi)/k_B T},
\end{equation}
where $n_0$ is the peak density, $k_B$ is the Maxwell-Boltzmann constant, $T$ is the temperature, and $U(r,\phi)$ is the confining potential.
The confining potential is given by Equation~(\ref{potential}), where $I(r) = |u^1_0(r)|^2$ for a pure \lgmode{0}{1} mode.
Since the intensity of this mode is azimuthally symmetric, it does not fit our data.
Tracing a path azimuthally around the absorption image, we note an oscillatory behavior in the atomic density distribution with a frequency of 3~Hz.
We add a term to the model for the potential that oscillates with the same frequency.
Our phenomenological model of the potential becomes:
\begin{equation}
	\label{potential_azimuthal}
	U_{\mathrm{LG}}(r,\phi) = -a \frac{2 r^2}{w^2} e^{-r^2/w^2}\left(1 + b\sin(3\phi+\phi_0)\right),
\end{equation}
where $b$ and $\phi_0$ are fitting parameters that give the fractional size and location of the azimuthal asymmetries in the intensity profile.  
The parameter $a = \hbar\Gamma^2 I_0/4 I_{\mathrm{sat}}\Delta k_B T$ is also a fitting parameter, where $I_0 = P/\pi w^2$ is the peak intensity of the trapping laser, and $P$ is the power of the trapping laser.
To simplify our model, we approximate the potential as a simple harmonic oscillator.
Expanding the potential to the second order about the equilibrium position $r = w/\sqrt{2}$, the potential becomes:  
\begin{equation}
	\label{harmonic_azimuthal}
	U_{\mathrm{sho}}(r,\phi) = \left(\frac{4a}{e w^2}\left(r - \frac{w}{\sqrt{2}}\right)^2 - \frac{a}{e}\right)\left(1+b\sin(3\phi+\phi_0)\right)+\frac{a}{e} (1+ b),
\end{equation}
where the last term defines the zero of the potential to be at the minimum.
Combining Equation~(\ref{harmonic_azimuthal}) with Equation~(\ref{density}) gives the density distribution function:
\begin{equation}
	\label{density_azimuthal}
	n(r,\phi)=n_0\exp\left\{-\left(\frac{4a}{e w^2}\left(r-\frac{w}{\sqrt{2}}\right)^2-\frac{a}{e}\right)\left(1+b\sin(3\phi+\phi_0)\right)-\frac{a}{e}(1+b)\right\}.
\end{equation}
\begin{figure}
	\centering
	\includegraphics[width=13.5cm]{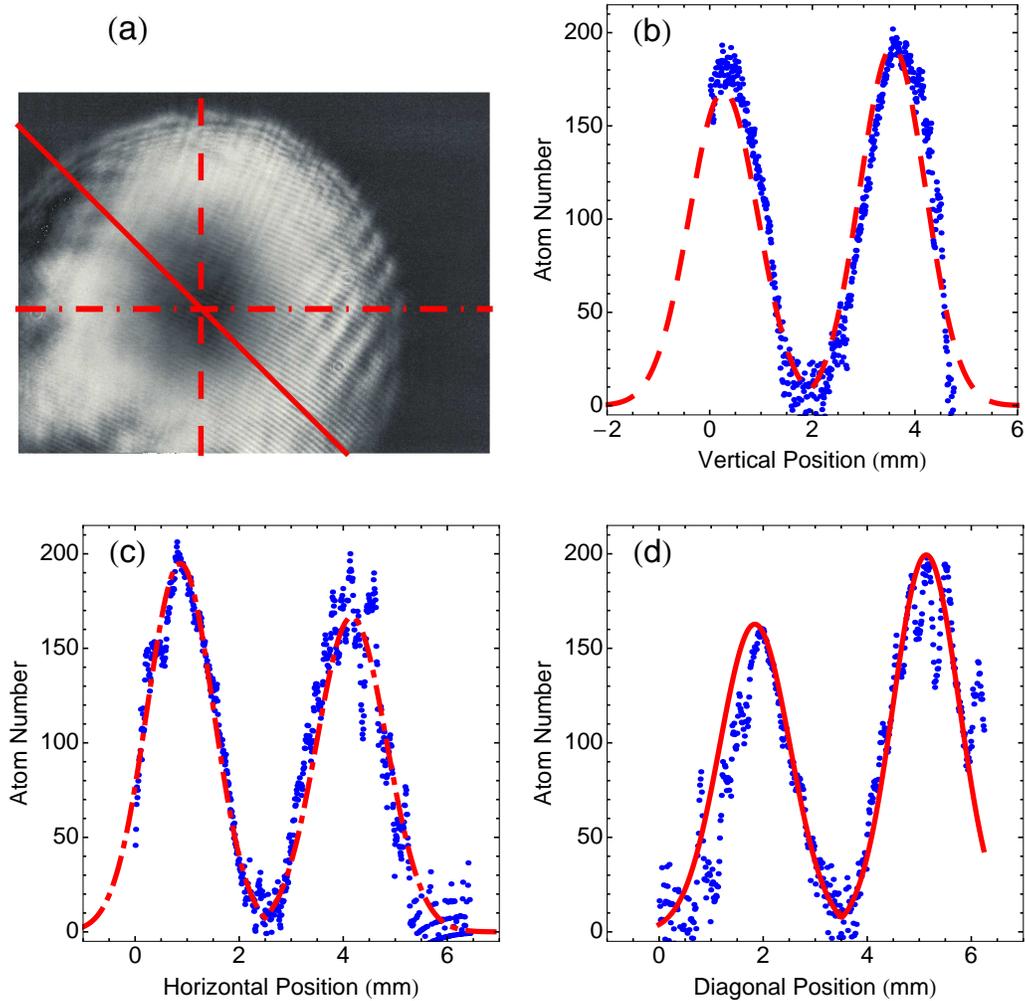}
	\caption{
		(Color online) (a) A 2D CCD image of the data shown in \ref{redwoGaussian}~(a).
		Each cross sections in (b)-(d) is indicated in the CCD image (a).
		The dashed red curve shows the vertical cross section, the dot-dashed red curve shows the horizontal cross section, and the solid red curve shows the diagonal cross section.
		The blue line is the data.
		For each cross section, $\phi$ is fixed and $r$ is varied.
		(d) Shows the largest azimuthal density difference.  
		\label{crosssection}   
	}
\end{figure}
\indent Equation~(\ref{density_azimuthal}) is fit to the data in Figure~\ref{redwoGaussian}~(a), using \textit{Mathematica}{\tiny\texttrademark}, where $a$, $b$, $w$, and $\phi_0$ are fitting parameters.
Figure~\ref{redwoGaussian}~(b) shows our best fit of our model for the density distribution.
From the fit, we determine that the laser intensity varies by 13\%.
From our known laser power and the fitting results, we can also conclude that the temperature of the two-dimensionally confined atoms is 13~$\mu$K.\\ 
\indent Figure~\ref{crosssection}~(a) is a 2D CCD image of the data shown in Figure~\ref{redwoGaussian}~(a). 
Each cross section for Figures~\ref{crosssection}~(b)-(d) is indicated in the CCD image, Figure~\ref{crosssection}~(a).
Figures~\ref{crosssection}~(b)-(d) show 1D cross sections of this data for fixed azimuthal angle $\phi$.
The blue points in each of the figures represent the absorption data along the cross section.
The red dashed, dot-dashed, and solid lines are the theoretical fits from the model in Equation~(\ref{density_azimuthal}).
Figure~\ref{crosssection}~(d) shows the largest azimuthal density difference. \\
\indent Sixty cross sections of the data in Figure~\ref{redwoGaussian}~(a), in which the radial coordinate $r$ is fixed but $0\le\phi\le2\pi$, were taken in 40~$\mu$m steps.  
Figure~\ref{azimuthal}~(a) is the same as Figure~\ref{crosssection}~(a), where the cross section at the radial antinode is indicated.
Figure~\ref{azimuthal}~(b) shows an average of seven cross-sections centered at the peak density.
The blue points represent the absorption data along the cross section.
The red line is the theoretical fit from the model in Equation~(\ref{density_azimuthal}).
The $\sin3\phi$ variation in density is clear in the data, and consistent with effects qualitatively observed in the intensity distribution of the \lgmode{0}{1} laser beam~\cite{kst02}.
For cross sections interior and exterior to the peak density, the $\sin 3\phi$ azimuthal trend is difficult to resolve due to the noise in the data including large variations due to interference in the probe laser from multiple reflections\footnote{A \textit{Mathematica}{\tiny\texttrademark}~notebook containing the data and all 60 cross-sections can be found online at~\cite{notebook}}. \\
\begin{figure}
	\centering
	\includegraphics[width=8cm]{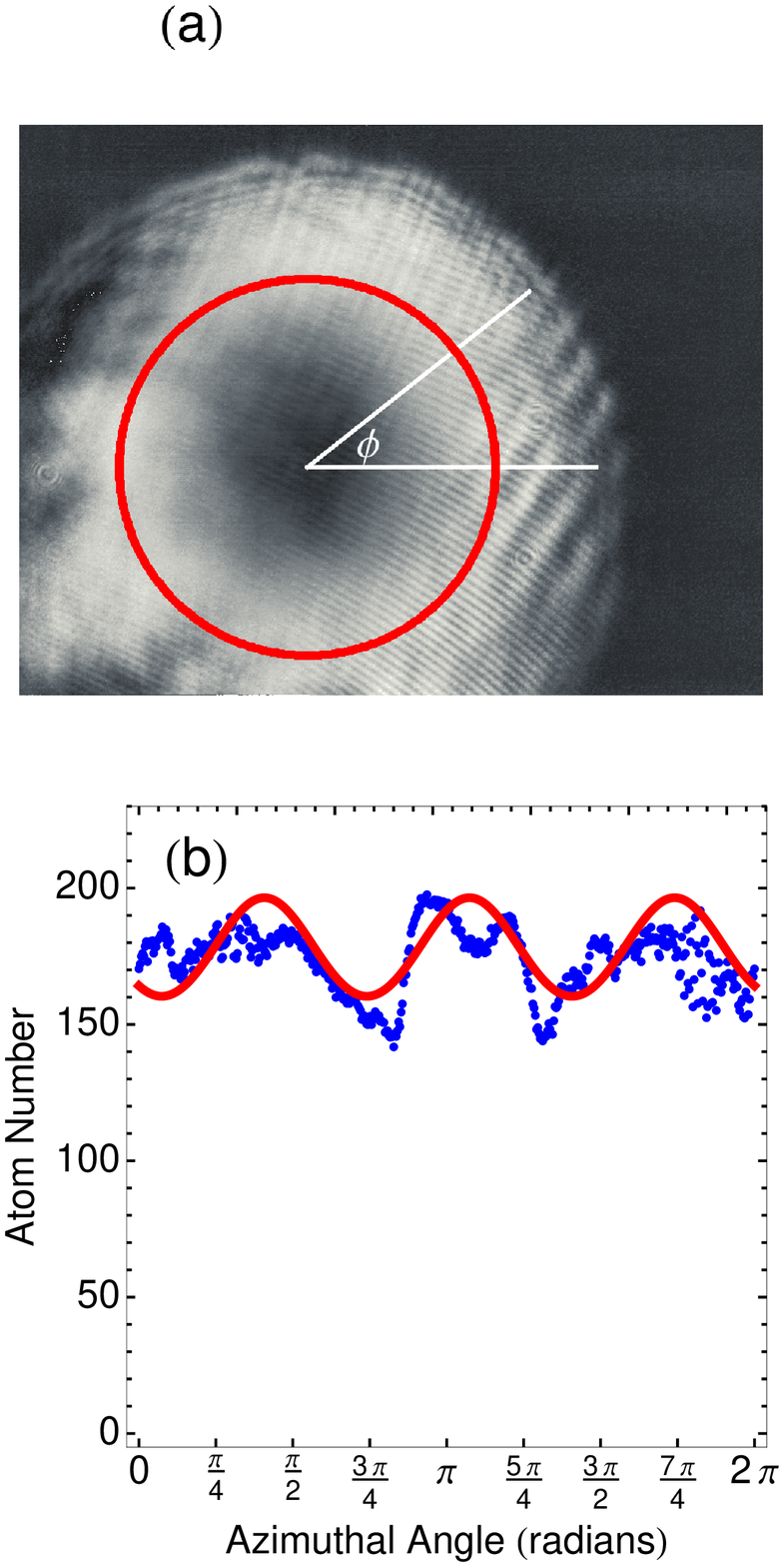}
	\caption{
		(Color online) (a) A 2D CCD image of the data shown in \ref{redwoGaussian}~(a).
		The cross section is indicated in the CCD image by the solid red curve.
		(b) The blue points are the average of seven cross section separated by 40~$\mu$m, centered on the peak density.
		For each cross section, $r$ is fixed and $0\le\phi\le2\pi$.
		The red curve is a fit of the data using Equation~\ref{density_azimuthal}, showing the $\sin3\phi$ variations in the density.
		\label{azimuthal}
	}
\end{figure}
\indent We successfully demonstrated 2D confinement of atoms from a MOT for both a blue-detuned \lgmode{0}{1} laser mode in which the atoms were confined in the center node of the laser beam and for a red detuned \lgmode{0}{1} laser mode in which the atoms were confined in the high-intensity ring. 
Asymmetries in the azimuthal intensity profile of the trapping laser appear as density inhomogeneities.
From our model of the density distribution, we determined that the confining potential fluctuates by 13\% along the azimuthal direction. 
We did not make any attempt to optimise azimuthal symmetry in this study.
From previous work~\cite{kst02}, these variations can be eliminated with better alignment through the diffractive optics.
However, this work also indicates that a systematic study of the purposeful misalignment of the diffractive optics in warranted to create new confinement potentials such as ferris wheel traps \cite{flp07}.\\
\indent High purity (symmetric) \lgmode{p}{\ell} modes are important for degenerate gas applications, gyroscopes, and vortex mater-wave creation, stability, and interferometry.
We have shown that analysis of the atomic distribution is a method to analyze the symmetry of the \lgmode{p}{\ell} transverse mode \textit{in situ} for systems where regular analysis of the full, high-intensity trapping beam is not feasible.
Also, purposeful creation of periodic, azimuthal intensity variations in the \lgmode{p}{\ell} modes may provide multiple traps for the toroidal geometries, increasing the range of experiments accessible to \lgmode{p}{\ell} beams made by diffractive optics.\\

This work was supported by The Research Corporation, Digital Optics Corporation, and the University of Oklahoma.
%
%
%--------------------------------------------------------------------------
%	BibTex References
%--------------------------------------------------------------------------
%
%
\bibliographystyle{elsarticle-num}
\bibliography{bibliography_lg_confinement}
\newpage
%--------------------------------------------------------------------------
%
%
\end{document}